\def\nk{n_{\rm b}}

\def\dT#1{\frac{\mathrm{d} #1}{\mathrm{d}T}}
\def\dTT#1{\frac{\mathrm{d} ^{2}#1}{\mathrm{d}T^{2}}}

\def\Di#1#2{\frac{\mathrm{D} #1}{\mathrm{d}#2}}
\def\mb#1{\mathbf{#1}}
\def\beq{\begin{equation}}
\def\eeq{\end{equation}}

\def\rfr#1{Equation\,(\ref{#1})}

\def\dertt#1#2{\frac{{{\textrm{d}^2}}{#1}}{{{\textrm{d}}}{#2}^2}}
\def\eqi{\begin{equation}}
\def\eqf{\end{equation}}
\def\eqia{\begin{eqnarray}}
\def\eqfa{\end{eqnarray}}

\def\rp#1#2{{#1\over#2}}

\def\bds#1{\boldsymbol{#1}}
\def\ton#1{\left(#1\right)}
\def\qua#1{\left[#1\right]}
\def\grf#1{\left\{#1\right\}}

\documentclass{ws-ijmpd}
\usepackage[super,compress]{cite}
\usepackage{hyperref}
\usepackage{booktabs,color}
\usepackage[utf8]{inputenc}
\begin{document}

\markboth{Lorenzo Iorio, Matteo Luca Ruggiero}
{Perturbations of the orbital elements due to the magnetic-like part of  the field of a plane gravitational wave}

\catchline{}{}{}{}{}
%
%%%%%%%%%%%%%%%%%%%%%%%%%%%%%%%%%%%%%%%%%%%%%%%%%%%%%%%%%%%%%%%%%%%%

\title{Perturbations of the orbital elements due to the magnetic-like part of  the field of a plane gravitational wave}

\author{LORENZO IORIO\footnote{Permanent address for correspondence: Viale Unit\`{a} di Italia 68, 70125, Bari (BA).}}

\address{Ministero dell'Istruzione, dell'Universit\`{a} e della Ricerca (M.I.U.R.)\\
lorenzo.iorio@libero.it}

\author{MATTEO LUCA RUGGIERO}

\address{Politecnico di Torino, Torino (Italy) and INFN, Laboratori Nazionari di Legnaro, Legnaro  (Italy)\\
matteo.ruggiero@polito.it}

\maketitle

\begin{history}
\received{Day Month Year}
\revised{Day Month Year}
\end{history}

\begin{abstract}
We focus on the secular changes of the orbital elements of a planet in the solar system, determined by the magnetic-like part of a gravitational wave field. Using Fermi coordinates we show that the total force acting on a test particle is made of two contributions: a gravito-electric one and a gravito-magnetic one. While the electric-like force has been thoroughly discussed in the past, the effect of the gravito-magnetic force, which depends on the velocity of the test particle,  has not been considered yet. We  obtain approximated results to some orders in the orbital eccentricity and show that these effects are much smaller than the corresponding gravito-electric ones.
\end{abstract}

\keywords{general relativity and gravitation;  celestial mechanics; gravitational waves}

\ccode{PACS numbers:}

%\tableofcontents

\section{Introduction}

Einstein's theory of gravity, General Relativity (GR), naturally predicts the existence of gravitational waves (see Ref. \refcite{flanagan2005basics} for a review). Indeed, already in 1905, Poincar\'e \cite{poincare1906dynamique} discussing Lorentz transformations, suggested that gravitational interactions could not be instantaneous, as in Newtonian theory, but should propagate at the speed of light.  Actually, it is possible to show (see e.g. Ref. \refcite{schutz1984gravitational}) that the basic properties of the gravitational waves (GWs) can be obtained by combining Newtonian gravity with the retardation effects due to the finite size of the speed of light. 
Einstein\cite{Einstein:1916cc,Einstein:1918btx}  made the first calculations defining the characteristics of the quadrupole formula for the emission of GWs. According to this formula, in order to have substantial emission of GWs,   huge masses moving at relativistic velocities are needed. These sources are  far away from the Earth and the weakness of the gravitational interaction  entails that what can be actually measured by our devices are very small perturbations or ripples in the fabric of space-time. Indirect evidence for gravitational waves was first obtained by observing the orbital decay of  the pulsar B1913+16\cite{Hulse:1974eb,Weisberg:2004hi}. It took 100 years from Einstein's first calculation to have, in 2016, the first direct detection of GWs ì\cite{abbott2016observation}, which was the beginning of \textit{gravitational waves astronomy}. 
More in general, the study and the detection of GWs could give information not only on GR, but also on the reliability of extended theories of gravity\cite{Capozziello:2011et,capozziello2010beyond}.

As far as we know (see e.g. Ref. \refcite{prince2009promise}), the spectrum of gravitational waves spreads over some 18 orders of magnitude, thus encompassing astrophysical and cosmological events on very different scales. In particular, while high frequency waves ($10-10^{4}$ Hz) are in the target of terrestrial detectors, there is no hope to detect on the Earth low-frequency  ($10^{-5} -1$ Hz) and very low frequency waves ($10^{-9} -10^{-7}$ Hz): in fact the ground motions and the coupling to fluctuations in the local mass distribution limit the sensitivity up to 1 Hz (see Ref. \refcite{flanagan2005basics}). While GWs in the low frequency regime are ideal target of the LISA space mission\cite{Tinto_2005,amaro2012low},  as for the very low frequency regime possible sources could be unresolved coalescing massive binary black holes or binaries
that are either too massive to radiate in the LISA band, or else are inspiralling towards the
LISA band\cite{flanagan2005basics}. 

Bertotti\cite{1973ApL....14...51B} first suggested to use gravitationally bound systems to detect secular effects due to a GW background: a Newtonian binary system consisting of point masses orbiting each other on Keplerian ellipses is acted upon by a gravitational wave, which causes a perturbation of the orbital elements.  There have been several subsequent approaches (see e.g.  Refs. \refcite{1975SvA....19..270R,1979ApJ...233..685T,1987SvA....31...76I,Hui:2012yp}) that are  thoroughly discussed by Iorio \cite{Iorio:2011eu}. Mashhoon\cite{mashhoon1977tidal} first developed a description of the tidal effects associated to the passage of a gravitational wave using Fermi coordinates,  then in Ref. \refcite{mashhoon1978tidal} he studied the interaction of a weak gravitational wave and a Newtonian binary system.

Fermi coordinates\cite{1922RendL..31...21F,manasse1963fermi,ni1978inertial} are defined around the world-line of an observer and are the coordinates he  would naturally use to make space and time measurements. It is possible to show\cite{Ruggiero_2020} that using Fermi coordinates the effects of a plane gravitational wave can be described by gravito-electromagnetic fields\cite{Ruggiero:2002hz,Mashhoon:2003ax}: in other words, the wave field is equivalent to the action of a gravito-electric and a gravito-magnetic field, that are transverse to the propagation direction and orthogonal to each other. In particular, while current detectors, such LIGO and VIRGO, or future ones, such as LISA, reveal the interaction of test masses with the gravito-electric component of the wave, there are also gravito-magnetic interactions that could be used to detect the effect of the gravitational wave on moving masses and spinning particles\cite{biniortolan2017,Ruggiero_2020,spinreso}.  {We point out that the effect of the angular momentum of gravitational waves on the rotation of inertial frames is studied in Refs. \refcite{PhysRevD.85.124003,Barker_2017}, while the theoretical possibility of measuring by a Sagnac detector the gravito-magnetic effects caused by a gravitational wave  is investigated by Frauendiener\cite{Frauendiener_2020}.}

In this paper we study the effect of the gravito-magnetic part of a gravitational wave, in the very low frequency regime, on the long-term variations of all the six standard Keplerian orbital elements of a solar system planet. This work extends to the magnetic-like components of the wave the study of the interaction with a Newtonian binary system, according to an approach similar to  Ref. \refcite{Iorio:2011eu}, where it is also pointed out that  very low frequency waves are important to get information on astrophysics of galaxies and black holes, early Universe cosmology and on the  extension of the standard model.

The paper is organised as follows: in Section \ref{sec:gmf} we briefly review gravito-electromagnetic approach to the description of the GWs field, in the Fermi frame; the perturbations of the orbital elements are worked out in Section \ref{sec:pert}, while conclusions are eventually drawn in Section \ref{sec:conc}.

%------------------------Section-------------------------
\section{Gravito-electromagnetic fields in the Fermi frame} \label{sec:gmf}
%------------------------Section-------------------------

In this Section we are going to briefly review the construction of the Fermi frame to describe the dynamics of test particles using the gravito-electromagnetic analogy. In order to define the Fermi frame, we use the approach described in Refs. \refcite{MTW,marzlin}. 

We consider a background spacetime describing the gravitational field where we use a set  of coordinates\footnote{Greek indices refer to spacetime coordinates, and assume the values $0,1,2,3$, while Latin indices refer to spatial coordinates and assume the values $1,2,3$, usually corresponding to the coordinates $x,y,z$.} $x^{\mu}$. Using these coordinates, the world-line $x^{\mu}(\tau)$ of a reference observer, as function of the proper time $\tau$, is determined by the following equation
\beq
\Di{x^{\mu}}{\tau}=\ddot x^{\mu}+\Gamma^{\mu}_{ \ \nu\sigma}\dot x ^{\nu} \dot x ^{\sigma} = a^{\mu}, \label{eq:eqmotion1}
\eeq
where $\mathrm D$ is the covariant derivative along the world-line, a dot means derivative with respect to $\tau$ and $a^{\mu}$ is the four-acceleration.  In the tangent space along the world-line $x^{\mu}(\tau)$ it is possible to define the orthonormal tetrad of the observer $e^{\mu}_{(\alpha)}(\tau)$, such that $e^{\mu}_{(0)}(\tau)$ is the unit vector tangent to his world-line and $e^{\mu}_{(i)}(\tau)$ (for i=1,2,3) are the spatial vectors  orthogonal to each other and, also, orthogonal to $e^{\mu}_{(0)}(\tau)$. The equation of motion of the tetrad is $\Di{e^{\mu}_{(\alpha)}}{\tau}=-\Omega^{\mu\nu} e_{\nu (\alpha)}$ where $\Omega^{\mu\nu}=a^{\mu} \dot x ^{\nu} - a^{\nu}\dot x^{\mu}+\dot x_{\alpha}\Omega_{\beta}\epsilon^{\alpha\beta\mu\nu}$. In the latter equation $\Omega^{\alpha}$ is the four-rotation of the tetrad. In particular, we notice that for a geodesic ($a^{\mu}=0$) and non rotating ($\Omega^{\alpha}=0$) tetrad we have  $\Omega^{\mu\nu}=0$: consequently, in this case the tetrad is parallel transported. If $\Omega=0$ and $a^{\mu} \neq 0$, the tetrad is Fermi-Walker transported. We remember that  Fermi-Walker transport enables to define, in a natural way,  a non rotating moving frame for an accelerated observer \cite{MTW}. 
That being said, Fermi coordinates are defined as follows: the observer along the congruence measures time intervals according to the proper time, so the time coordinate is defined by $T=\tau$; the spatial coordinates $X,Y,Z$ are defined by space-like geodesics, with unit tangent vectors $n^{\mu}$, whose components, with respect to the orthonormal tetrad are $n^{(i)}=n_{(i)}=n_{\mu}e^{\mu}_{(i)}(\tau)$, and $n^{(0)}=0$. The reference observer's frame equipped with Fermi coordinates is the \textit{Fermi frame}. 
Fermi coordinates in the neighbourhood of the world-line of an observer in accelerated motion  with rotating tetrads are studied in Refs. \refcite{Ni:1978di,Li:1979bz,marzlin}.

Fermi coordinates prove useful to develop a gravito-electromganetic analogy. Indeed, as discussed by Mashhoon\cite{Mashhoon:2003ax}, the spacetime element in Fermi coordinates in the vicinity of the observer's world-line can be recast  in terms of the gravito-electromagntic  potentials $(\Phi, \mb A)$  
\begin{equation} 
ds^2=-\left(1-2\frac{\Phi}{c^2}\right)c^{2}dT^2-\frac{4}{c}({\mb 
A}\cdot d{\mb
X})dt+\delta_{ij}dX^idX^j \, , \label{eq:mmetric2}
\end{equation}
 and this peculiarity allows us to apply the corresponding formalism\footnote{Here and henceforth $\mb X$ is the position vector in the Fermi frame.}.  Indeed, as shown by \cite{Ruggiero_2020}, it is possible to distinguish between the \textit{inertial}  and the \textit{curvature} contributions to the gravito-electromagnetic potentials: the former depend on the inertial features of the Fermi frame (i.e. the observer's acceleration and tetrad rotation), while the latter  on the Riemann curvature tensor. Here, we are concerned with the curvature effects only, since we refer our results to an inertial frame: for instance, if we consider planetary motion in the solar system, its origin is at the position of the Sun. Consequently, the   \textit{gravito-electric} (GE) potential $\Phi=\Phi (T, {\mb X})$ is
\beq
\Phi (T, {\mb X})=-\frac{1}{2}R_{0i0j}(T )X^iX^j \label{eq:defPhiG}
\eeq
and the components of the \textit{gravito-magnetic}  potential (GM) $\mb A=\mb A (T ,{\mb X})$ turn out to be
\beq
A^{}_{i}(T ,{\mb X})=\frac{1}{3}R_{0jik}(T )X^jX^k. \label{eq:defAG}
\eeq
In close analogy to electromagnetism, the gravito-electric and gravito-magnetic fields $\mb E$ and  $\mb B$ are defined in terms of the potentials by
\begin{equation} {\mb E}=-\nabla \Phi 
-\frac{1}{c}\frac{\partial}{\partial T}\left( \frac{1}{2}{\mb
A}\right),
\quad {\mb B}=\nabla \times {\mb A}. \label{eq:defEB1}
\end{equation}
Using the definitions (\ref{eq:defPhiG}) and (\ref{eq:defAG}) we obtain (up to linear order in $|X^{i}|$) the following components
\beq
E^{C}_i(T ,{\mb X})=c^{2}R_{0i0j}(T) X^j, \label{eq:defEIEG}
\eeq
and 
\beq
B^{C}_i(T ,{\mb R})=-\frac{c^{2}}{2}\epsilon_{ijk}R^{jk}_{\;\;\;\; 0l}(T )X^l. \label{eq:defBIBG}
\eeq

The electromagnetic analogy is also useful to describe the motion of free test particles: namely, the geodesic  equation of the space-time metric (\ref{eq:mmetric2})   can be written in the form of a Lorentz-like force equation \cite{Mashhoon:2003ax}
\beq
m\dTT{\mb X}=q_E{\mb E}+q_B\frac{\mb V}{c}\times {\mb B},  \label{eq:llorenz1}
\eeq
up to linear order in the particle velocity $ \mb V=\dT{\mb X}$ (which is indeed the \textit{relative velocity} with respect to the reference mass at the origin of the frame).  In the Lorentz-like force equation, $q_{E}=-m$ is the GE charge, and $q_{B}=-2m$ is the GM  one (the minus sign takes into account that the gravitational force is always attractive). As a consequence, the Lorentz-like force equation becomes
\beq
m\dTT{\mb X}=-m{\mb
E}-2m\frac{\mb V}{c}\times {\mb B}. \label{eq:lorentz0}
\eeq

In the following Section we are going to work out the effect of the gravito-magnetic field of the wave on a moving test particle. For the sake of simplicity, we consider  plane monochromatic circularly polarised GWs propagating with frequency $\nu_\mathrm{g}$ along the direction $Z$. Moreover, in what follows we suppose that the extension of the reference frame is much smallar than the wavelength, so that we may neglect the spatial variation of the wave field:  consequently the components of the Riemann tensor are evaluated at the origin of our frame. Accordingly, as shown in \cite{Ruggiero_2020}, the corresponding gravito-magnetic field is

\begin{align}
B^{}_{Z}  &= 0,\\\nonumber \\
B^{}_{X}  &= -\frac{A\,\nu^2_\mathrm{g}}{2}\,\ton{- X\,\cos\chi + Y\,\sin\chi},\\ \nonumber \\
B^{}_{Y}  &= -\frac{A\,\nu^2_\mathrm{g}}{2}\,\ton{X\,\sin\chi + Y\,\cos\chi}, \label{eq:defBxyz12}
\end{align}
with $\chi\doteq\nu_\mathrm{g} T$.

Hence, the gravito-magnetic force acting upon a point-like mass $m$ is obtained by
\begin{equation}
m\,\dertt{\bds X}{T} = - 2\,m\,\rp{\bds V}{c}\bds\times \bds B. \label{eq:lorentz}
\eqf
Note that, being $B$ dimensionally an acceleration, the scaling parameter $A$ is dimensionless.

\section{Perturbations due to the gravito-magnetic field of the wave } \label{sec:pert}

To study the effect of the magnetic-like part of the gravitational field of the wave on a planet of the solar system we proceed as follows. We consider a Fermi coordinate system, whose origin is located at the Sun position; accordingly, we treat the planet as a test particle moving around the Sun.\footnote{More in general, studying a binary system, the origin of the Fermi frame should be located at the center of mass of the system, and what we would consider the relative position of the two masses (see e.g. Ref, \refcite{chicone1996gravitational}). The approximation considered is reasonable for most of the bodies of the solar system.  } 
Let us assume that the wave's frequency $\nu_\mathrm{g}$ is much lower than the orbital one $\nk$, so that the averaged rates of change of the orbital elements of the test particle can be analytically calculated with the Gauss perturbing equations \cite{Sof89,1991ercm.book.....B,2003ASSL..293.....B,2005ormo.book.....R,2011rcms.book.....K,2014grav.book.....P,SoffelHan19} by keeping $\chi= \mathrm{const}$ in the integration over the true anomaly $f$ from $0$ to $2\Pi$ of the latter.
The radial, transverse and normal components of the acceleration of \rfr{eq:lorentz} turn out to be
\begin{align}
A_\mathrm{r} \label{Ar} & = -\frac{A\,a^2\,\sqrt{1-e^2}\,\nk\,\nu^2_\mathrm{g}\,\sin I}{c}\,\qua{\cos I\,\cos\ton{\chi + 2 \Omega}\,\sin u + \cos u\,\sin\ton{\chi + 2 \Omega}}, \\ \nonumber \\
A_\mathrm{t} \label{At} & = \frac{A\,e\,a^2\,\sqrt{1-e^2}\,\nk\,\nu^2_\mathrm{g}\,\sin I\,\sin f}{c\,\ton{1 + e\,\cos f}}\,\qua{\cos I\,\cos\ton{\chi + 2 \Omega}\,\sin u +
\cos u\,\sin\ton{\chi + 2 \Omega}}, \\ \nonumber \\
A_\mathrm{n} \label{An} & = \frac{A\,a^2\,\sqrt{1-e^2}\,\nk\,\nu^2_\mathrm{g}}{16\,c\,\ton{1 + e\,\cos f}}\,\ton{2\,e\,\cos\ton{f - \chi - 2 \Omega} -
e\,\cos\ton{f - 2 I - \chi - 2 \Omega} - \right.\\ \nonumber \\
\nonumber & -\left. 2\,\cos\ton{2 I - \chi - 2 \Omega} + 2\,\qua{2 - \cos 2 I + 2\,\ton{3 + \cos 2 I}\,\cos 2u} \cos\ton{\chi + 2 \Omega} + \right. \\ \nonumber \\
\nonumber & + \left. 4\,\cos I\,(\sin I - 4\,\sin 2u)\,\sin\ton{\chi +2 \Omega} + \right.\\ \nonumber \\
\nonumber & + \left. e\,\grf{-\cos\ton{f + 2 I - \chi - 2 \Omega} + 2\,\qua{(3 + \cos 2 I)\,\cos\ton{f - \chi + 2 \omega - 2 \Omega} + \right.\right.\right. \\ \nonumber \\
\nonumber & + \left.\left.\left. \cos\ton{f + \chi + 2 \Omega} + 3\,\cos\ton{f + \chi + 2\varpi} - 8\,\cos I\,\sin\ton{f + 2 \omega} \sin\ton{\chi + 2 \Omega} - \right.\right.\right. \\ \nonumber \\
&-\left.\left.\left. 2\,\cos 2 I\,\sin \omega\,\sin\ton{f + \chi + \omega + 2 \Omega}}}},
\end{align}
where $a$ is the semimajor axis, $e$ is the eccentricity, $I$ is the inclination to the reference $\grf{x,\,y}$ plane, $\Omega$ is the longitude of the ascending node, $\omega$ is the argument of pericentre, $u\doteq \omega+ f$ is the argument of latitude, and $\varpi\doteq\Omega+\omega$ is the longitude of pericentre.
An exact calculation in $e$ turns out to be exceedingly cumbersome; thus, we will only show approximate results to some orders in $e$.

To the order of $\mathcal{O}\ton{e^{-1}}$, nonzero secular rates occur for the argument of pericentre and the mean anomaly at epoch $\eta$. They are
\begin{align}
\dot\omega &= \frac{a\,A\,\nu^2_\mathrm{g}\,\sin I\ton{\cos I\,\cos\ton{\chi + 2\Omega}\sin\omega + \cos\omega\sin\ton{\chi + 2\Omega}  }}{2\,c\,e}, \\ \nonumber \\
\dot\eta &= -\dot\omega.
\end{align}

To the order zero in $e$, only the eccentricity experiences a long-term rate of change
\begin{equation}
\dot e = \frac{a\,A\,\nu^2_\mathrm{g}\,\sin I\,\ton{-\cos I\cos\omega\cos\ton{\chi + 2\Omega} + \sin\omega\sin\ton{\chi + 2\Omega}}}{2\,c}. \label{eq:dtoe}
\end{equation}

To the order of $\mathcal{O}\ton{e}$, the inclination $I$, the node $\Omega$, the pericentre $\omega$ and the mean anomaly at epoch $\eta$ undergo secular precessions given by
\begin{align}
\dot I \label{inc} \nonumber & = \frac{e\,a\,A\,\nu^2_\mathrm{g}}{16\,c}\,\qua{-2\,\cos\ton{I - \chi - \omega - 2 \Omega} + \cos\ton{2 I - \chi - \omega - 2 \Omega} +
 2\,\cos\ton{I - \chi + \omega - 2 \Omega} + \right.\\ \nonumber \\
\nonumber & +\left.  \cos\ton{2 I - \chi + \omega - 2 \Omega} +
 2\,\cos\omega\,\ton{-6\,\cos\ton{\chi + 2 \Omega} + \cos\ton{2 I + \chi + 2 \Omega}} + \right.\\ \nonumber \\
 &+\left. 4\,\sin\omega\,\sin\ton{I + \chi + 2 \Omega}}, \\ \nonumber \\
\dot \Omega  \nonumber & = \frac{e\,a\,A\,\nu^2_\mathrm{g}\,\csc I}{8\,c}\,\grf{-\sin\ton{I - \chi - \omega - 2 \Omega} -
 \sin\ton{2 I - \chi - \omega - 2 \Omega} -
 \sin\ton{I - \chi + \omega - 2 \Omega} + \right.\\ \nonumber \\
\nonumber & +\left.  \sin\ton{2 I - \chi + \omega - 2 \Omega} +
 \sin\ton{I + \chi - \omega + 2 \Omega} -
 \sin\ton{2 I + \chi - \omega + 2 \Omega} +\right.\\ \nonumber \\
&+\left.  \sin\ton{I + \chi + \omega + 2 \Omega} +
 \sin\ton{2 I + \chi + \omega + 2 \Omega}}, \\ \nonumber \\
\dot \omega \label{om} & = \frac{e\,a\,A\,\nu^2_\mathrm{g}}{4\,c}\,\qua{-2\,\cos\ton{\chi + 2 \Omega}\,\cot I\,\sin\omega + \ton{-3 +\cos 2I}\,\cos\omega\,\csc I\,\sin\ton{\chi + 2 \Omega}}, \\ \nonumber \\
\dot \eta \label{et} & = -\frac{7\,e\,a\,A\,\nu^2_\mathrm{g}\,\sin I}{4\,c}\,\ton{\cos I\,\cos\ton{\chi + 2 \Omega} \sin\omega + \cos\omega \sin\ton{\chi + 2 \Omega}} 
\end{align}

If, instead, it is $\nu_\mathrm{g}\gg\nk$, it turns out that there are no net orbital shifts averaged over a wave's full cycle.

In the previous results, we made no assumptions on the orbital inclination $I$; we notice that, for $I=0$, i.e. the incidence of the gravitational wave is normal to the orbital plane, we see that there is no secular variation of the eccentricity, and the same holds true for the mean anomaly at epoch.

\section{Conclusions}\label{sec:conc}

%The paper somewhat too suddenly ends. Since the formulas are not entirely short, it would be good to try to conclude in words what is the effect of waves on planetary orbits, I mean both of the electric and of the magnetic components. In a long term, could signals from astrophysical sources somehow affect the Solar system? Could they contribute to make the system at least very slightly chaotic? Does this imply anything for detection of waves?

In this paper we studied the averaged rates of change of the orbital elements of a test particle orbiting a central body, such as a planet orbiting the Sun, determined by the magnetic-like force due to the passage of a monochromatic plane gravitational wave. In doing so, we assumed that the wavelength is larger then the orbit size and, also, that the  frequency is much smaller than the orbital frequency of the test particle. 

To fix ideas,  in the solar system we remember that Mercury has the highest orbital frequency, $1.3 \times 10^{-7}$ Hz. Hence, in this case our results apply to  the very low frequency regime of gravitational waves. However, our results could be applied to other astrophysical situations, such as planets in extrasolar systems, where higher orbital frequencies are possible.

Since exact calculations are  complicated, in order to give an idea of the effects, we worked out approximated results to some orders in the orbital eccentricity $e$. Our results showed that all orbital elements, except the semi-major axis, undergo secular variations;  {in particular, we proved that to zero order in the eccentricity $e$, only the eccentricity itself experiences a long-term rate of change. This
 variation is a purely gravito-magnetic effect: in fact, as discussed in  Ref. \refcite{Iorio:2011eu}, if we consider the gravito-electric perturbation only, the variation of the eccentricity is proportional to the eccentricity itself, so that circular orbits do not change their shape. More in general, orbits with given eccentricities change their shape and  orientation due to the passage of the wave:  all variations determined by the gravito-magnetic effects are proportional to $\displaystyle \epsilon=\frac{aA\nu^2_\mathrm{g}}{c}$, which is a very small quantity, depending on the wave amplitude $A$, the orbital frequency $\nu_\mathrm{g}$ and semimajor axis $a$, and the speed of light $c$. However, even if these gravito-magnetic effects are in principle present, it is reasonably expected that they are negligible with respect to the gravito-electric ones. This can be explained by inspecting Eq. (\ref{eq:lorentz}): in fact, while both  the gravito-electric and magnetic fields have the same order of magnitude, the gravito-magnetic force is reduced, with respect to the gravito-electric one, by a factor $\displaystyle \frac{|\mb V|}{c}$, where the magnitude of the orbital speed $|\mb V|$ is much smaller than the speed of light.} 

 {In conclusion, the aim of this  paper was to focused on  the gravito-magnetic effects provoked by the passage of a plane gravitational wave; as discussed by one of us\cite{Iorio:2011eu}, there are also gravito-electric effects which, according to the above discussion, have a larger impact. However, the separation of the two effects is purely academic, since a point mass experiences a total force in the form (\ref{eq:llorenz1}). As a consequence,  to make predictions on  the impact of a gravitational wave on orbital elements we need to consider the whole perturbing acceleration determined by the passage of the wave: this will be  the subject of a future work.  These predictions could be used to make a comparison with planetary data available from the solar system  which can, in principle, give constraints on all unknowns wave parameters.  This formalism could be applied not only in the solar system, but to arbitrary gravitationally bound two-body systems, such as extrasolar systems, and with suitable modifications, it can be used to test models of gravity alternative to General Relativity, thus adding new perspectives to the field of gravitational waves astronomy.}

\bibliographystyle{ws-ijmpd}

%\bibliography{2PN}{}

%
\end{document}